\documentclass[%
 reprint,
 superscriptaddress,
 amsmath,amssymb,
 aps,
 pre,
 onecolumn,
floatfix,
]{revtex4-2}

\usepackage{graphicx}
\usepackage{dcolumn}
\usepackage{bm}
\usepackage{graphicx} 

\usepackage{hyperref}
\usepackage{subcaption}
\usepackage{bbm}
\usepackage{adjustbox}

\input{preamble}

\newcommand{\der}[0]{\text{d}}
\newcommand{\invt}[0]{\beta^{-1}}
\newcommand{\ttot}[0]{t_{\rm tot}}
\newcommand{\was}[0]{\mathcal{W}_2}
\newcommand{\ctrl}[1]{#1^{(c)}}

\graphicspath{{.}}

\begin{document}

\title{Universal energy-speed-accuracy trade-offs in driven nonequilibrium systems}
\author{J\'{e}r\'{e}mie Klinger}
\affiliation{Department of Chemistry, Stanford University, Stanford CA, USA 94305}
\author{Grant M. Rotskoff}
\affiliation{Department of Chemistry, Stanford University, Stanford CA, USA 94305}
\affiliation{Institute for Computational and Mathematical Engineering, Stanford University, Stanford CA, USA 94305}
\date{\today}

\begin{abstract}
    The connection between measure theoretic optimal transport and dissipative nonequilibrium dynamics provides a language for quantifying nonequilibrium control costs, leading to a collection of ``thermodynamic speed limits'',
    which rely on the assumption that the target probability distribution is perfectly realized. This is almost never the case in experiments or numerical simulations, so here, we address the situation in which the external controller is imperfect.
    We obtain a 
    lower bound for the dissipated work in generic nonequilibrium control problems that 1) is asymptotically tight and 2) matches the thermodynamic speed limit in the case of optimal driving.
    Along with analytically solvable examples, we refine this imperfect driving notion to systems in which the controlled degrees of freedom are slow relative to the nonequilibrium relaxation rate, and identify independent energy contributions from fast and slow degrees of freedom. Furthermore, we develop a strategy for optimizing minimally dissipative protocols based on optimal transport flow matching, a generative machine learning technique.
    This latter approach ensures the scalability of both the theoretical and computational framework we put forth. 
    Crucially, we demonstrate that we can compute the terms in our bound numerically using efficient algorithms from the computational optimal transport literature and that the protocols we learn saturate the bound. 
\end{abstract}

\maketitle

\section{Introduction}
Trade-offs involving dissipation, control, and fluctuations are a unifying feature among the few universal results that constrain nonequilibrium dynamics. 
While fluctuation theorems~\cite{crooks_entropy_1999, lebowitz_gallavotticohen-type_1999, kurchan_fluctuation_1998}, thermodynamic uncertainty relations~\cite{barato_thermodynamic_2015, gingrich_dissipation_2016}, and thermodynamic speed limits~\cite{aurell_refined_2012} all have broad applicability, for many physical systems the bounds computed from these fundamental relations are too weak for inference of the dissipation rate~\cite{gingrich_inferring_2017} or are difficult to compute directly in many-body systems~\cite{chennakesavalu_probing_2021}.

Recent efforts to adopt a purely geometric perspective on nonequilibrium dynamics are creating an opportunity for new analytical and computational approaches that more precisely constrain and quantify dissipation, even far from equilibrium. 
Foundational work in the partial differential equations literature~\cite{jordan_variational_1998} revealed a deep connection between dissipation and optimal transport theory; in particular, the relaxation to equilibrium described by a Fokker-Planck equation follows a Wasserstein geodesic and the geometry of the geodesic relates to relaxational entropy production.
Aurell et al.~\cite{aurell_optimal_2011, aurell_refined_2012} made the connection explicit in the language of stochastic thermodynamics and showed that the total entropy production of a driven nonequilibrium transformation can be directly expressed in terms of the Monge-Kantorovich optimal transport distance.
This relation, in turn, leads to thermodynamic speed limits~\cite{nakazato_geometrical_2021, VanVu2022}.

Unfortunately, the difficulty of solving high-dimensional optimal transport problems, both analytically and numerically, limits the applicability of this conceptually elegant approach.
Currently, it is both difficult to identify external control protocols that drive a system \emph{accurately} towards a fixed target distribution and also \emph{minimize dissipation}, an important goal in many settings~\cite{jarzynski_targeted_2002, rotskoff_geometric_2017, neal_annealed_2001, vaikuntanathan_escorted_2011}.  
Existing strategies for minimum dissipation protocol optimization either require a linear response approximation~\cite{sivak_thermodynamic_2012, crooks_measuring_2007, rotskoff_geometric_2017}, necessitate exact knowledge of the optimal transport mapping~\cite{chennakesavalu_unified_2023}, or require exhaustive simulations that realize trial protocols~\cite{gingrich_near-optimal_2016, engel_optimal_2023}. 

In this article, we derive two relations between optimal transport and dissipation that extends this framework to imperfect control.
Because optimal transport problems in high-dimension are analytically intractable, we develop a machine learning approach to identify low dissipation protocols, without the significant computational expense of carrying out dynamical simulations.
Our algorithm is closely related to recently developed generative machine learning techniques~\cite{tong_improving_2023, liu_flow_2022, albergo_building_2023, lipman_flow_2022} that provide new and exciting opportunities to solve high-dimensional inference problems on which classical numerical methods fail~\cite{engel_optimal_2023, karniadakis_physics-informed_2021, boffi_deep_2023}.
We demonstrate that this framework yields control forces that accurately drive many-particle systems towards arbitrary targets. 
Because the solutions obtained by any computational approach will be necessarily approximate, we also refine the theoretical formulation of the thermodynamic speed limits to account for the imperfect control protocols.
The most significant contribution of our work is an argument that optimal transport methods are indispensable, even in the case of imperfect control---a fact not widely appreciated in the nonequilibrium dynamics literature~\cite{VanVu2022, van_vu_topological_2023, chennakesavalu_probing_2021}. 
Furthermore, the dissipation from the controlled and uncontrolled degrees of freedom can be unambiguously delineated in some cases, as we show below. 
In addition, we derive a straightforward, but strikingly general bound \eqref{eq:mainbound} by formalizing the notion of accurate control.
Together, these results articulate universal trade-offs among energy dissipation, speed of driving, and accuracy of control.

\section{Optimal transport and dissipation}
Throughout, we consider finite-time nonequilibrium optimal control problems in which we seek to drive a system with initial probability density $\rho_0$ to a target density $\rho_*$ at time $\ttot$. 
We consider a classical system with coordinates $\xb \in \Omega \subset \RR^d$, and assume that trajectories of this system evolve according to the overdamped Langevin equation
\begin{equation}
\label{eq:langevin}
    \der \Xb_t = b(\Xb_t, t) \der t + \sqrt{2 \beta^{-1}} \der \Wb_t, \quad \Xb_0 = \xb.
\end{equation}
Here the drift $b:\RR^d\to\RR^d$ generically contains both conservative and non-conservative forcing terms, $\beta^{-1} = k_{\rm b} T$, and $\Wb_t$ is a $d$-dimensional Wiener process.
Following the conventions of stochastic thermodynamics~\cite{Seifert2008}, we write the single trajectory entropy production as
\begin{equation}
\label{eq:defentropy}
\begin{alignedat}{1}
    \Delta &S_{\rm tot}[\Xb_{[0,\ttot]}] =\Delta S_{\rm sys} - \beta Q[\Xb_{[0,\ttot]}]\\
    &= \log \frac{\rho_{\ttot}(\Xb_{\ttot})}{\rho_0(\Xb_0)} - \beta \int_0^{\ttot} b(\Xb_s,s)\circ\der \Xb_s.
    \end{alignedat}
\end{equation}
The average system entropy change $\Delta S_{\rm sys}$ is a state function only involving the initial density $\rho_0$ and the final density $\rho_{\ttot}$.
The second term in \eqref{eq:defentropy} is a time-symmetric Stratonovich integral, which is conventionally interpreted to be proportional to the heat flow along the trajectory, $Q[\Xb_{[0,t]}]$.
Remarkably, as shown by Aurell et al.~\cite{aurell_refined_2012}, taking an expectation over trajectories, the average total entropy production can be bounded by an optimal transport distance between the initial and final distributions
\begin{equation}
 \Delta S_{\rm tot} \geq \beta \ttot^{-1} \mathcal{W}_2^2(\rho_0, \rho_{*}),
 \label{eq:speedlimit}
\end{equation}
where the Monge-Kantorovitch distance $\mathcal{W}_2$ is defined as \cite{villani_optimal_2009}
\begin{equation}
    \mathcal{W}_2^2(\rho_0, \rho_*) = \underset{T}{\inf}\int_{\Omega} |\xb-T(\xb)|^2\rho_0(\xb)\der \xb
    \label{eq:l2}
\end{equation}
over all transport plans $T$ mapping $\rho_0$ onto $\rho_*$, $i.e$ such that $T\#\rho_0 = \rho_*$.
This formulation immediately leads to a number of useful insights; first, it clearly establishes a lower bound on the entropy production rate for systems driven to a target state in a finite time, a result explored in several recent works~\cite{nakazato_geometrical_2021, VanVu2022, dechant_geometric_2022, yoshimura_housekeeping_2023}.
Secondly, the dissipation is minimal when the optimal transport map $T\equiv T_*$ is known exactly.

While the optimal transport formulation is a mathematically elegant and conceptually clear framing that yields a finite-time refinement of the Second Law, the crucial assumption underlying the speed limit~\eqref{eq:speedlimit}---that the final distribution is perfectly realized---cannot be imposed easily for arbitrary target densities $\rho_*$.
Indeed, in many systems that we may want to control, the realizable transport plans will be limited by the set of external couplings available.
Of course, this means that the minimizer of \eqref{eq:l2} may not be reached.

\begin{figure}[ht]
    \centering
    \includegraphics[width = 0.5\textwidth]{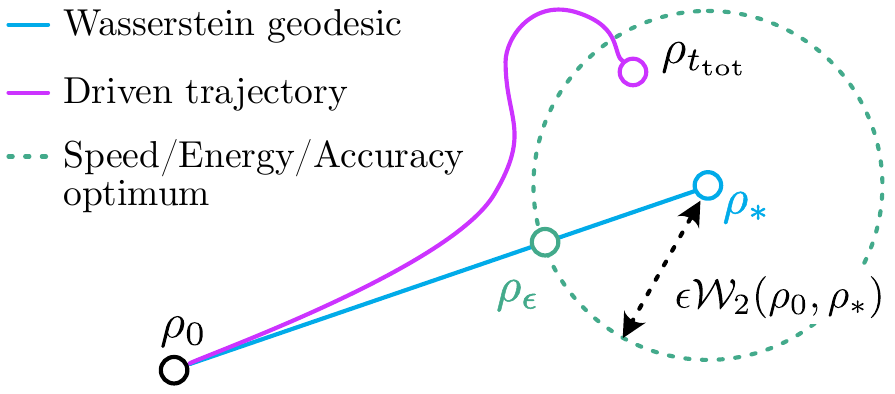}
    \caption{The minimum dissipation required to produce an $\epsilon$-accurate distribution $\rho_{\epsilon}$ is derived based on purely geometric criteria. This argument leads to a generic energy-speed-accuracy trade-off~\eqref{eq:mainbound}. Constant speed geodesics in Wasserstein space are depicted as straight lines in the schematic. }
    \label{fig:schematics}
\end{figure}

Failing to reach the target distribution $\rho_*$ alters the constraint on the total dissipation; trivially, if we make no change to the distribution, there is no excess entropy production from the control. 
A more productive setting to consider is one in which we closely realize the target distribution, but do not satisfy exactly the final time constraint.
Suppose we set a target ``accuracy'' which we measure in the optimal transport distance: we say that a distribution $\rho$ is $\epsilon$-accurate if $\was(\rho, \rho_*) \leq \epsilon \was(\rho_0, \rho_*)$, as illustrated in Fig.~\ref{fig:schematics}.
We now ask about the cost of getting close to the target distribution; in other words, we address how the lower bound on the dissipation is influenced by the accuracy parameter $\epsilon$. 

The optimal controller drives the system along a Wasserstein geodesic at all times, as illustrated in Fig.~\ref{fig:schematics}.
Because $\mathcal{W}_2$ is a proper distance metric on the space of probability densities, it is straightforward to see via the triangle inequality that the minimal dissipation to produce an $\epsilon$-accurate distribution $\rho$ at time $t_{\rm tot}$ must result from following the geodesic between $\rho_0$ and $\rho_*$.   
A physically realizable external coupling that we use to drive a given system may not follow this optimal trajectory at every point in time.
As a result, we immediately obtain the energy speed accuracy (ESA) bound 
\begin{equation}
\begin{aligned}
 \label{eq:mainbound}
    \ttot \Delta S_{\rm tot} \geq \beta(1-\epsilon)^2 \mathcal{W}_2^2(\rho_{0}, \rho_{*}), 
\end{aligned}
\end{equation}
where $\epsilon = \mathcal{W}_2(\rho_{\ttot}, \rho_{*})/\mathcal{W}_2(\rho_{0}, \rho_{*})$ and the right-hand side is a purely geometric quantity depending on only the distances between the source, target, and $\epsilon$-accurate density. 
In other words the minimum entropy production associated to an inaccurate controller scales in proportion to the discrepancy measured in Wasserstein space.

\begin{figure}[ht!]
    \centering
    \includegraphics[height = 0.4\textwidth]{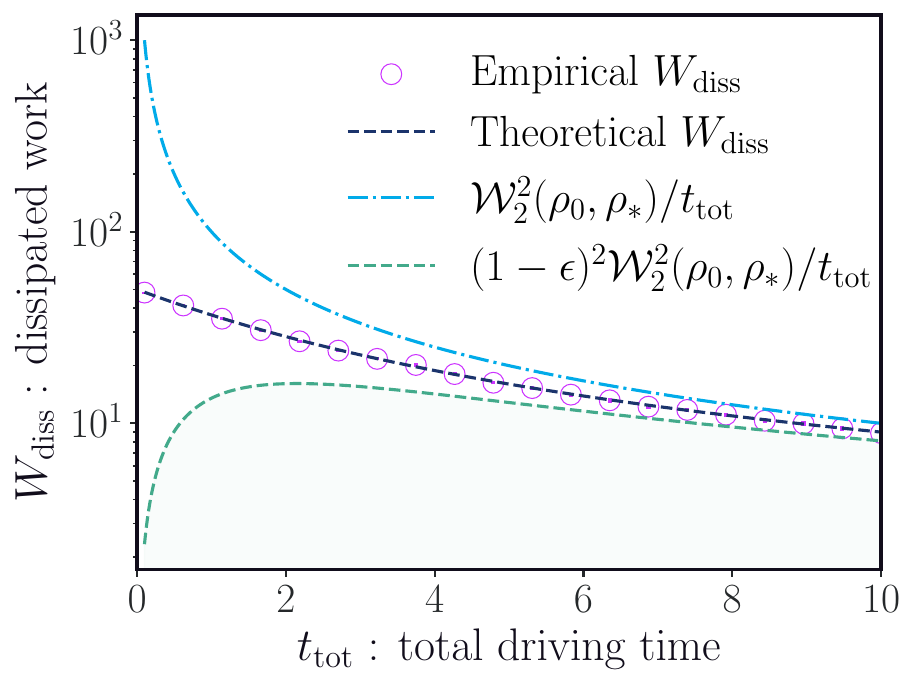}
    \caption{The thermodynamic speed limit does not lower bound $W_{\rm diss} \equiv \beta^{-1}\Delta S_{\rm tot}$ for inaccurate driving, while the ESA bound does. We plot the total entropy production along the linear driving protocol from $\lambda_0 = 0$ to $\lambda_{*} = 10.0$ for a 1$d$ OU processes with constant stiffness parameter and inverse temperature and $\beta = k = 1.0$. The dissipated work $W_{\rm diss}$ can be computed exactly and is shown as a dashed black line, in agreement with numerical results. Note that the speed limit is not expected to hold, specifically because the instantaneous distribution is different from $\rho_*$.
    }
    \label{fig:1d-OU}
\end{figure}

We illustrate this discrepancy on the analytically tractable example of a Brownian particle linearly driven by a harmonic trap. The resulting dynamics are that of a Ornstein-Uhlenbeck process, for which all Gaussian instantaneous distributions are known exactly. As a result, all terms in the ESA bound \eqref{eq:mainbound} can be exactly computed (and given in Appendix A) and compared with the empirical dissipation on Fig. \ref{fig:1d-OU}.
We emphasize that the energy-speed-accuracy bound~\eqref{eq:mainbound} is only as tight as the classical speed limit, provided that the measurement of $\mathcal{W}_2$ is performed between the initial distribution and the realized distribution $\rho_{\ttot},$ but it may be easier to measure this distance to an idealized target, as in the bound.

\section{High-dimensional control with flow-matching}
In high-dimensional systems, exact solutions to the optimal transport problem are generally unknown.
As a result, designing minimally dissipative control forces 
becomes challenging.
In the following, we develop an approach to low dissipation control based on rectified flow models~\cite{albergo_building_2023, lipman_flow_2022, liu_flow_2022}.

Indeed, recent extensions of continuous normalizing flows \cite{liu_flow_2022} and Optimal Transport Flow Matching (OTFM) suggest that minimizing the $\was$ for the transport map that defines the generative model aids optimization and model performance for image generation tasks~\cite{pooladian_multisample_2023, tong_improving_2023}. 
This approach naturally fits the paradigm of stochastic thermodynamics and, 
with an appropriate formulation, we extend this framework to provide an algorithmic toolkit for the design of minimally dissipative controllers, relevant even for high-dimensional physical systems. 

\subsection{Learning minimally dissipative control forces}
Broadly speaking, flow matching~\cite{chen_neural_2018,Onken2021, lipman_flow_2022, liu_flow_2022} is a generative modeling technique that parameterizes a flow field $u_t(x)$ such that the solution $\psi_t(x)$ to the ODE 
\begin{equation}
\label{eq:ot1b}
    \der \psi_t(x) = u_t(\psi_t(x))\der t
\end{equation}
with initial condition $\psi_0(x)=x \sim \rho_0$, transports $\rho_0$ to $\rho_*$ in time $\ttot$, i.e. $\psi_{\ttot}(x) \sim \rho_*$. 
For brevity, we set $\ttot = 1$ in the following. 
A natural strategy \cite{liu_flow_2022} is to seek trajectories with an instantaneous distribution that linearly interpolates between source and target distributions, i.e., we want the solution of the ODE~\eqref{eq:ot1b} to be equal in law to 
\begin{equation}
    \textrm{Prob}[\psi_t(x_0)] \equiv \textrm{Prob}[x_t] \quad \mathrm{ where }\quad x_t = t x_0 + (1 - t)x_1, \qquad x_0, x_1 \sim \rho_0 \times \rho_*,
\label{eq:ot2b}
\end{equation}
that is, $x_0$ and $x_1$ are respectively sampled according to $\rho_0$ and $\rho_*$ and $x_t$ is the linear interpolant connecting them. 
In this case, the initial and final time constraints are enforced by design.
By taking a time derivative of the desired interpolation $\eqref{eq:ot2b}$, one obtains a simulation-free objective for the flow field $u_t(x)$
\begin{equation}
    \mathcal{L}_{\rm FM}\left[u\right] = \mathbb{E}\left[||(x_1-x_0) - u_t(x_t)||^2\right],
\label{eq:ot3b}
\end{equation}
where $x_0$ and $x_1$ are sampled from $\rho_0$ and $\rho_*$, and $x_t = t x_0 + (1-t)x_1$, with $t \sim \mathcal{U}([0,1])$.
Importantly, the evaluation of the objective \eqref{eq:ot3b} does not require the potentially costly dynamical integration of the ODE~\eqref{eq:ot1b}.
While minimizing the objective \eqref{eq:ot3b} guarantees that $\psi_1(x_0) \sim \rho_*$, the instantaneous distribution $\rho_t$  of $\psi_t(x_0)$ need not follow the Wasserstein geodesic between source and target distribution.
Optimal transport Flow Matching (OTFM) refines the objective~\eqref{eq:ot3b} by requiring that the pair $(x_0, x_1)$ be jointly sampled from the Optimal Transport coupling $\pi^{\rm OT}(x_0, x_1) = \rho_0(x_0)\delta(x_1-T^*(x_0))$, where $T^*$ is the optimal transport plan~\eqref{eq:l2}. 
By definition of the Wasserstein geodesic (referred to as the displacement interpolant in Chap. 5 of \cite{ villani_topics_2003})
\begin{equation}
\rho_t = ((1-t)\text{Id} + t T^*)\# \rho_0
\label{eq:ot4b}
\end{equation}
the minimizer of the associated OTFM objective \eqref{eq:ot3b} drives the system $\psi_t(x_0)$ along this very geodesic, thus realizing the optimal dynamical control from source to target.
Exploiting the deep connection between optimal transport and stochastic thermodynamics, we now tailor OTFM to nonequilibrium control problems. 
In particular, temperature fluctuations resulting in deviations from the OT pathway need to be controlled.

To motivate the general approach, we consider a particle in contact with a thermal bath, whose dynamics are given by the one-dimensional overdamped Langevin equation, as in \eqref{eq:langevin}
%
%
and we seek a control force $b^{\rm c}$ to drive the system from $\rho_0$ to $\rho_*$ with minimum dissipation.  The corresponding Fokker-Planck equation is given by
\begin{equation}
    \label{eq:ot6}
    \partial_t \rho_t = -\nabla\left[(b^{\rm c}-\invt \nabla \log \rho_t)\rho_t\right], \hspace{10pt}\rho_{t=0}=\rho_0
\end{equation}
and the least dissipative protocol drives $\rho_t$ along the constant speed Wasserstein geodesic between source and target. 
To account for deviations induced by thermal fluctuations, we decompose the optimal controller as $b^{\rm c} = u^\theta + \invt s^\theta$. While the driving force $u^\theta$ minimizes the conditional OTFM objective \eqref{eq:ot3b}, the contribution $s^\theta$ cancels out the instantaneous ``score'' $\nabla \log \rho_t$, maintaining the system on its geodesic course.
The score associated to SDE-based generative models \cite{hu_diffusion_2017,song_score-based_2022} has received increasing attention in the past few years and can be approximated in a number of ways \cite{lipman_flow_2022, albergo_building_2023}. In particular, the score is the unique minimizer of the objective function \cite{hyvarinen_estimation_2005}
\begin{equation}
    \label{eq:ot7}
    \mathcal{L}_{\rm score}\left[s\right] = \mathbb{E}\left[2\nabla \cdot s_t(x_t) + ||s_t(x_t)||^2\right]
\end{equation}
where, again, $x_t = t x_0 + (1-t)x_1$ with $(x_0, x_1) \sim \pi^{\rm OT}$.
Consequently, both $u^\theta$ and $s^\theta$ can be trained concurrently to learn the thermodynamically optimal controller $b^*$.

This methodology relies on a number of approximations. First, the quality of the controller depends on the convergence of the learning procedure. 
More importantly, accurate sampling from the optimal coupling $\pi^{\rm OT}$ is not necessarily feasible for high-dimensional distributions, and one needs to resort to approximate methods, such as the Sinkhorn algorithm \cite{cuturi_sinkhorn_2013} or direct empirical estimates of optimal transport distance.
Fortunately, these approaches still provide scalable and controlled approximations of the optimal coupling~\cite{mena_statistical_2019}.
Moreover, we emphasize that even if potentially inaccurate, the control $b^*$ allows for dissipation and accuracy gains, whose interplay is quantified by the general ESA bound~\eqref{eq:mainbound}.

\begin{figure}[h!]
    \centering
    \hspace{-30pt}
    \includegraphics[height = 0.4\textwidth]{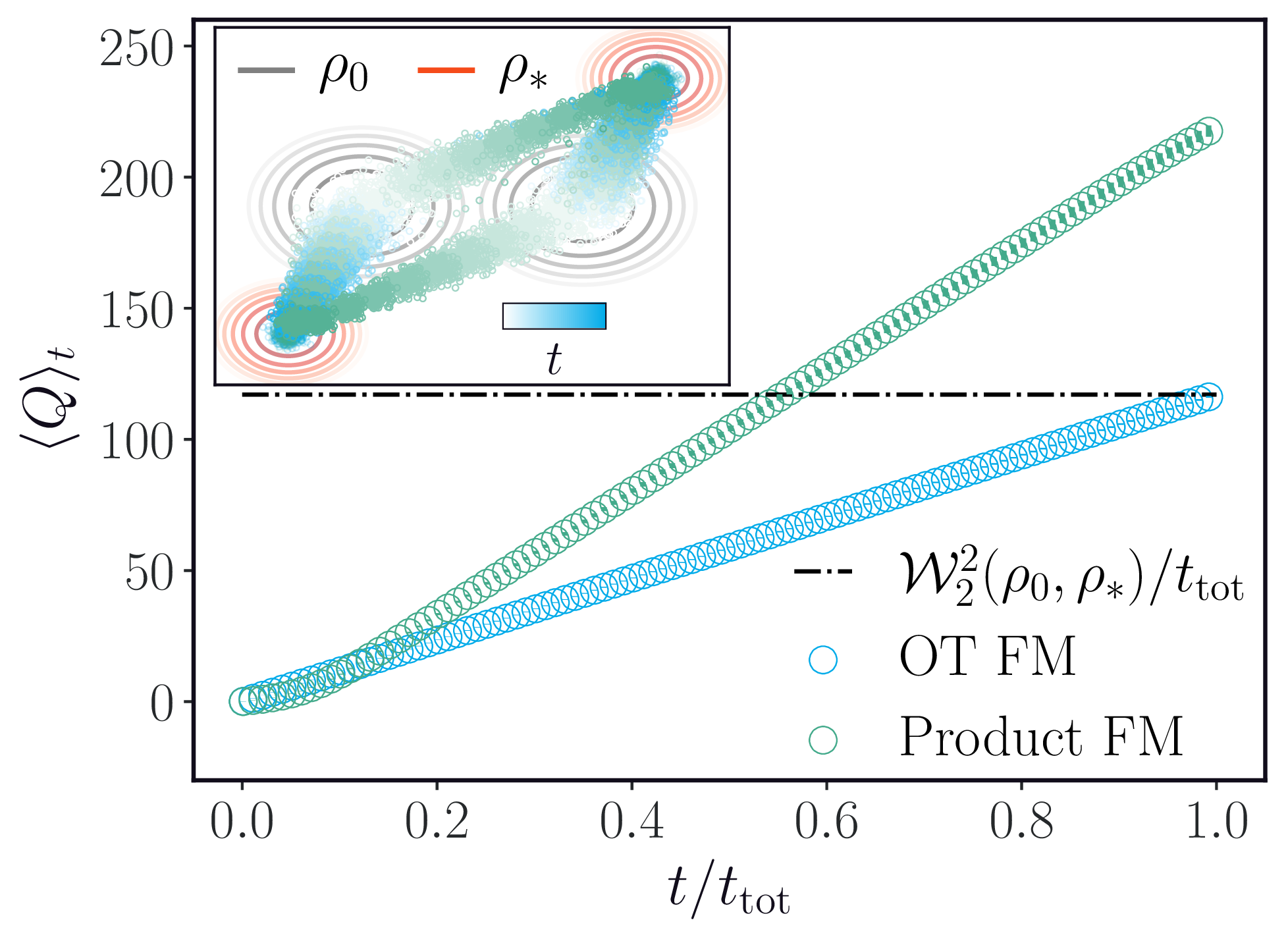}
    \captionsetup{
    }
    \caption{Empirical dissipated heat along the bimodal to bimodal controlled dynamics \eqref{eq:langevin} with $\beta = 1.0$. The highly dissipative nature of the trajectories associated to $b^\times$ stems from the apparent excess displacement. \textbf{(inset)} Source and target distribution isodensity lines, and dynamical samples generated from the learned controllers $b^\times$ and $b^{\rm OT}$. Both controllers display satisfying end point performance, but the space-time trajectories of the driven samples are very different.}
    \label{fig:bimodal2}
\end{figure}

The OTFM approach dramatically lowers the dissipative cost of transport for systems that are fully controllable, as shown in Fig. \ref{fig:bimodal2}. 
To illustrate saturation of the bound, in this example transport task between distinct $2d$ bimodal distributions, we assume that we can independently couple to each degree of freedom, an assumption we relax in what follows. 
Details regarding the source and target distribution, as well as the training procedure, are provided in Appendix C.
Crucially, we compare two optimizations of the Flow Matching objective \eqref{eq:ot3b}. The product controller $b^{\times}$ is obtained by training on product pairs $(x_0, x_1) \sim \rho_0(x_0)\rho_*(x_1)$, while the optimal controller $b^{\rm OT}$ is trained on empirical optimal coupling pairs $(x_0, x_1) \sim \pi^{\rm OT}(x_0, x_1)$. The drastically different instantaneous distributions and dissipation rates emphasize the significance of attending to the optimal transport pathway. 
Note that while the minimally dissipative controller $b^{\rm OT}$ is trained on an empirical estimate of the source/target optimal coupling, it still provides near perfect driving.

The approach we employ is scalable by design, allowing straightforward application to complex systems of many interacting particles.
In Appendix D, we apply this technique to drive a cloud of quorum sensing active particles between two metastable states.
While this calculation again assumes arbitrarily precise control, it highlights both the saturation of the bound and the efficacy of numerical optimal transport for even complicated physical systems.

\subsection{Optimal control with weak coupling---timescale separation}
For many systems, external control is limited---the applied force only elicits response on a small number of degrees of freedom, and internal many-body interactions are uncontrolled.
In turn, inherent dissipative effects associated to these fast uncontrolled degrees of freedom contribute to the total excess heat along the driving.
For instance, active particles have a non-zero entropy production rate in any potential, but the shape of a global confining potential might be directly controllable by an external field.
In the following, we show that OTFM-type controllers are still optimal with respect to the dissipation incurred by an operator when the steady state entropy production occurs on a fast timescale.

We consider a class of systems comprising of a set of fast dissipative pairwise interacting degrees of freedom $\Xb_t$ coupled to a controlled degree of freedom $\Xbc_t$, which is driven between source and target distributions $\Xb_0^{\rm c} \sim \rho_0$ and $\Xb_{\rm t_{tot}}^{\rm c} \sim \rho_*$ via a controller $b^{\rm c}(\xb, t)$.
The equations of motions 
\begin{equation}
\begin{split}
&\der \Xbc_t = b^{\rm c}(\Xbc_t,t )\der t \\
&\der \Xb_t = \epsilon^{-1}F(\Xb_t - \Xbc_t)\der t+ \sqrt{2\invt\epsilon^{-1}}\der \Wb_t
\end{split}
\label{eq:ness1}
\end{equation}   
where the $\epsilon$ timescale accounts for the fast relaxation of the $\Xb_t$ degrees of freedom compared to the controller $\Xbc_t$.
In turn, the $\epsilon$ independent excess work is given by
\begin{equation}
W_{\rm ex} = \Delta S_{\rm sys} + \beta Q_{\ttot} = \Delta S_{\rm sys}  +\beta \int_0^{\ttot} F(\Xb_t - \Xbc_t)\circ\der \Xb_t.
\label{eq:ness2}
\end{equation}
The time scale separation hypothesis amounts to assuming that the instantaneous distribution of $\Xb_t$ is close to the stationary state associated to a fixed value of $\Xbc_t$. Using a two timescale expansion~\cite{rotskoff_geometric_2017} we show that the excess work can be factorized into controllable and uncontrollable parts.
We begin with the equilibrium case $F = -\nabla U$. Because the controller evolves according to a deterministic ODE, the work done along the trajectory can be conditioned on the initial value of the controller
\begin{equation}
    \mathbb{E}\left[W\right] =
    \int_\Omega \der \Xbc_0 \rho_0(\Xbc_0) \avg{W|\Xbc_0}
    = \int_\Omega \der \Xbc_0 \rho_0(\Xbc_0)\int_0^{\ttot} \der t \dot{\Xbc_t}\epsilon^{-1}\avg{\nabla U (\Xb_t - \Xbc_t) | \Xbc_0}_{\rho_t}
    \label{app:timescale2}
\end{equation}
where $\rho_t$ is implicitly the instantaneous distribution of $\Xb_t$ knowing $\Xbc_0$. Along a fixed $\{\Xbc_t\}_{t \geq 0}$ trajectory, we now perform a small $\epsilon$ expansion of the Fokker Planck equation associated to \eqref{eq:ness1}
\begin{equation}
    \partial_t \rho_t = -\epsilon^{-1}\nabla \left[-\nabla U(x-\Xbc_t)\rho_t - \beta^{-1}\nabla \rho_t \right].
    \label{app:timescale3}
\end{equation}
By expanding the instantaneous distribution around the stationary state
\begin{equation}
    \rho_t = \rho^{\rm ss}_t(1 + \epsilon \phi(\xb, t) + O(\epsilon^2))
    \label{app:timescale4}
\end{equation}
given by $\rho^{\rm ss}_t = e^{-\beta U(\xb - \Xbc_t)}$ (we absorb the normalization in $U$), it is seen \cite{rotskoff_geometric_2017} that the zeroth non vanishing order is given by
\begin{equation}
    \nabla \phi \nabla U + \beta^{-1} \Delta\phi = \beta \nabla U \dot{\Xbc_t}.
    \label{app:timescale5}
\end{equation}
Using the Feynman-Kac formula, the solution $\phi$ to this ODE is given by
\begin{equation}
    \phi(\xb, t) = \beta\dot{\Xbc_t}\mathbb{E}_x\left[\int_0^\infty \nabla U(\Tilde{\Xb}_\tau - \Xbc_t)\der \tau\right]
    \label{app:timescale6}
\end{equation}
where the expectation is taken with respect to a new stochastic process $\Tilde{X}_\tau$ whose dynamics obey
\begin{equation}
    \Tilde{X_0} \sim \rho^{\rm ss}_t,\hspace{10pt}\der \Tilde{\Xb_\tau} = -\nabla U(\Xb_\tau - \Xbc_t)\der \tau + \sqrt{2 \beta^{-1}}\der \Wb_\tau.
    \label{app:timescale7}
\end{equation}
Plugging the expansion~\eqref{app:timescale4} into~\eqref{app:timescale2} yields
\begin{equation}
    \begin{split}
        \avg{W | \Xbc_0} &= \int_0^{\ttot} \der t \dot{\Xbc_t}\left[\avg{\epsilon^{-1}\nabla U (\Xb_t - \Xbc_t) | \Xbc_0}_{\rho^{\rm ss}_t} + \int \rho^{\rm ss}_t(\xb)\phi(\xb ,t)\nabla U(\xb-\Xbc_t)\der \xb\right] \\
        &=\epsilon^{-1}\left[F(\Xbc_{\ttot}) - F(\Xbc_{0})\right] +  \beta\int_0^{\ttot} \der t \dot{\Xbc_t}\left(\int_0^\infty \der \tau \mathbb{E}\left[\nabla U (\Tilde{\Xb}_\tau - \Xbc_t)\nabla U (\Tilde{\Xb}_0 - \Xbc_t)\right]_{\rho^{\rm ss}_t}\right)\dot{\Xbc_t}
    \end{split}
    \label{app:timescale8}
\end{equation}
where $F(\Xbc)$ is the free energy of the fast degrees of freedom, for fixed controller value $\Xbc$. Because the system is translationnaly invariant, the free energy difference is zero, and the total excess work --- which is equal to the heat~--- is simply given by
\begin{equation}
    \avg{Q_{t_{\rm tot}}} = \beta\left(\int_0^\infty \der \tau \mathbb{E}\left[\nabla U (\Tilde{\Xb}_\tau)\nabla U (\Tilde{\Xb}_0)\right]_{\rho^{\rm ss}}\right)\int_\Omega \der \Xbc_0\ \rho_0(\Xbc_0)\int_0^{\ttot} \der t ||\dot{\Xbc_t}||^2.
    \label{app:timescale10}
\end{equation}
For any protocol $b^{\rm c}(\xb, t)$ driving the controlled degrees of freedom $\Xb^{\rm c}_t$ between $\rho_0$ and $\rho_*$, the OT pathway minimizes the right hand side of equation \eqref{app:timescale10}, such that the heat is lower bounded by
\begin{equation}
\avg{Q_{t_{\rm tot}}} \geq \beta\left(\int_0^\infty \der \tau \mathbb{E}\left[\nabla U (\Tilde{\Xb}_\tau)\nabla U (\Tilde{\Xb}_0)\right]_{\rho^{\rm ss}}\right) \frac{\was^2(\rho_0,\rho_*)}{\ttot}.
    \label{app:timescale11}
\end{equation}
The generic case of arbitrary forces $F$ proceeds similarly, and is detailed in Appendix E.
We show that the heat can be  decomposed into a controllable part and a housekeeping-like contribution, and the relevant dissipation associated to the driving $b^{\rm c}$ is lower bounded by
\begin{equation}
\avg{Q_{\ttot}} -\ttot \avg{\dot{Q}_{\rm hk}} \geq C \frac{\was^2(\rho_0, \rho_*)}{\ttot}
\label{eq:ness3}
\end{equation}
where $C$ is a system dependent constant, independent of the driving protocol $b^{\rm c}$, and the housekeeping heat~\cite{hatano_steady-state_2001} corresponds to the steady state dissipation at fixed $\Xbc$.
For equilibrium systems  $F\equiv -\nabla U$, $\dot{Q}_{\rm hk} = 0$ and the constant $C$ given by Eq. \eqref{app:timescale11}, and can be measured empirically. 

\begin{figure}[h!]
    \centering
    \hspace{-30pt}
    \includegraphics[height = 0.4\textwidth]{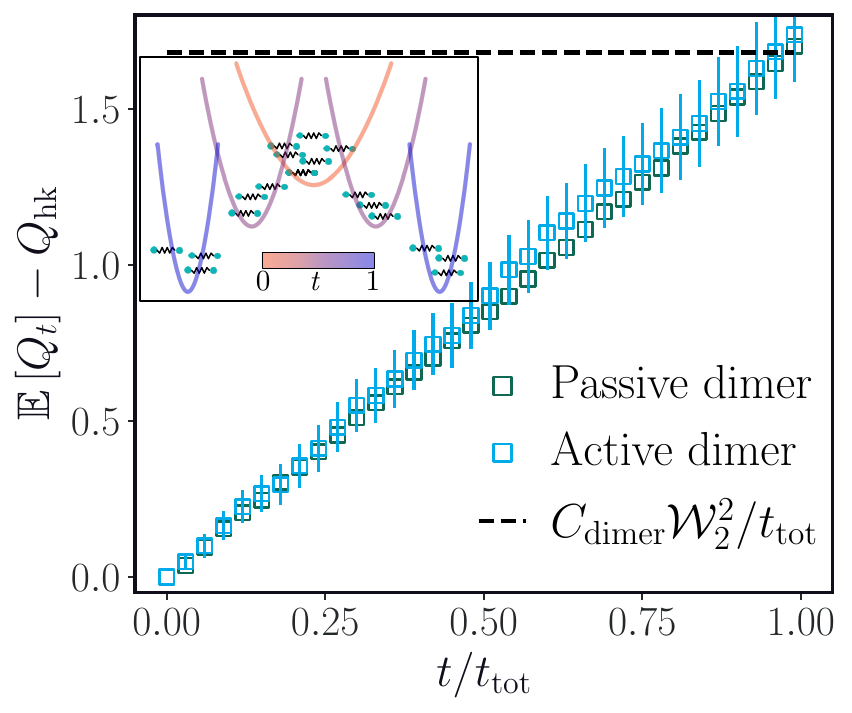}
    \captionsetup{
    }
    \caption{One-dimensional active/passive dimer optimally driven along the dynamics \eqref{eq:ness1} by a control parameter bridging two Gaussian mixtures. For a passive dimer, $Q_{\rm hk} = 0$, $C_{\rm dimer}$ is obtained numerically, and the bound is tight. The active dimer has additional QSAP-type dynamics yielding a non zero stationary dissipation, which is subtracted out. Details on the systems are given in Appendix F.}
    \label{fig:dimer-driving}
\end{figure}

This is our second main result, which emphasizes the importance of the OT strategy even for systems where only partial control can be implemented.
Specifically, in this slow driving regime, the bound is saturated when $b^{\rm c}$ is learned via the OTFM technique and $\Xbc_t$ is driven along the OT pathway.
Finally, we display on Fig.~\ref{fig:dimer-driving} the agreement between numerical simulations and the bound~\eqref{eq:ness3} for a passive/active dimer driven in a potential undergoing an optimal uni-modal to bimodal separation.

\section{Conclusion} Recent results that link control, dissipation, and speed~\cite{sivak_thermodynamic_2012, chennakesavalu_unified_2023, nakazato_geometrical_2021, dechant_geometric_2022} are built on deep mathematical connections between optimal transport theory and physical dynamics~\cite{jordan_variational_1998, aurell_optimal_2011}. 
Away from the idealized setting in which the exact optimal mapping can be identified \emph{and realized}, thermodynamic speed limits do not necessarily provide a meaningful constraint on dissipation.
Here, we demonstrate that an approach built on the geometry of optimal transport still supports precise lower bounds that not only relate dissipation and speed, but also quantify the thermodynamic cost of highly accurate control.
We show that the ESA bound tightly constrains the dissipation in analytically solvable models, where the optimal controller can be constructed exactly. 

Applications of optimal transport theory have been limited in scope for physical systems, perhaps due to the lack of scalable computational methods, but synergistic developments in generative machine learning are breaking down these computational barriers.
Our results, which use empirical estimates of Wasserstein distances together with optimal transport flow-matching provide strong evidence that high-dimensional controllers can be constructed which are nearly optimal thermodynamically. 

Here, our primary focus has been optimal control of diffusive classical systems.
Extensions of the thermodynamic connection between OT and stochastic thermodynamics to Markov jump processes~\cite{maas_gradient_2011} and open quantum systems~\cite{chen_matrix_2018, becker_quantum_2021, chennakesavalu_unified_2023} require further development and investigation. 
Furthermore, the provocative connection between our results, which relate dissipation, speed, and accuracy, with the thermodynamic uncertainty relations that indicate a universal trade-off between dissipation and the precision of fluctuations~\cite{barato_thermodynamic_2015, gingrich_dissipation_2016} merits further investigation, as well. 

\newpage
\appendix

\section{Linearly driven 1$d$ Ornstein–Uhlenbeck process}

In this section, we consider the excess dissipation $W_{\rm diss}\equiv \beta^{-1} \Delta S_{\rm tot}$ of a one-dimensional particle in a linearly displaced harmonic trap with fixed stiffness $k$ and inverse temperature $\beta$.~\cite{schmiedl_stochastic_2007, blickle_realization_2012}. 
The particle, initially at thermal equilibrium and distributed according to $\rho^{\rm eq}_{\lambda_0}\propto \text{exp}\left[-\beta k(x - \lambda_0)^2/2\right]$, is driven towards a target distribution $\rho_*^{\rm eq} \propto \text{exp}\left[-\beta k(x - \lambda_*)^2/2\right]$ along a finite time linear protocol 
\begin{equation}
\label{eq:OU-1}
    \lambda_t = \lambda_0 + (\lambda_* - \lambda_0)\frac{t}{\ttot}
\end{equation}
at constant temperature, before relaxing to the target distribution. In turn , we decompose the excess dissipation $\Delta S_{\rm tot}=\Delta S^{\rm drive}_{\rm tot}+\Delta S^{\rm relax}_{\rm tot}$ into contributions from driving and relaxation dynamics. 
Since $\Delta S^{\rm relax}_{\rm tot}\geq 0$, the ESA bound (6) from the main text yields a lower bound on the total entropy production. The equation of motion for the particle position during the driving phase is given by
\begin{equation}
\label{eq:OU1}
\der X_t = -k (X_t - \lambda_t) \der t+ \sqrt{2 \invt}\der W_t; \quad  X_0 \sim \rho_{\lambda_0}^{\rm eq}
\end{equation}
and the instantaneous distribution of $X_t$ is a Gaussian with mean and variance given by
\begin{equation}
\label{eq:OU2}
    \begin{alignedat}{1}
        &\avg{X_t} = \frac{\mu_0(1+k(\ttot-t))+\mu_*(kt-1)}{k\ttot}+\frac{e^{-kt}(\mu_*-\mu_0)}{k\ttot}\\
        &\avg{(X_t-\avg{X_t})^2} = \beta k.
    \end{alignedat}
\end{equation}
Because the variance of $\rho^{\rm eq}_{\lambda_0}$ coincides with that of the target, the free energy difference vanishes and one obtains the total entropy production
\begin{equation}
    \label{eq:OU3}
    \Delta S_{\rm tot} = \beta W_{\rm diss} = \beta \frac{\left[\ttot-k^{-1}(e^{-k\ttot}-1)\right](\lambda_0-\lambda_*)^2}{\ttot^2}.
\end{equation}
What is more, for Gaussian distributions, the optimal transport distance can be computed in closed form $\mathcal{W}_2^2(\rho_{\ttot}, \rho^{\rm eq}_{\lambda_*}) = (\avg{X_{\ttot}} - \lambda_*)^2$, from which one obtains the accuracy parameter $\epsilon$ exactly. In Fig. 2, we have $\lambda_0 = 0$, $\lambda_{*} = 10.0$ and $\beta = k = 1.0$.

\section{Optimal Transport Flow Matching - an alternative formulation}
\label{appSIscore}

In this section, we provide an alternative description of optimal transport flow matching (OTFM). In essence, this approach is very closely related to the one described in the main text, but, for consistency, we provide both discussions. Most of the following expressions are directly extracted from \cite{lipman_flow_2022} and reproduced here for the sake of completeness. 

Given source and target distributions $\rho_0$ and $\rho_*$, flow matching methods build upon the general Continuous Normalizing Flow (CNF) framework \cite{chen_neural_2018} which proposes to learn a flow field $u_t(x)$ such that the solution $\psi_t(x)$ to the ODE

\begin{equation}
\label{eqap:ot0}
    \der \psi_t(x) = u_t(\psi_t(x))\der t
\end{equation}
with initial condition $\psi_0(x)=x \sim \rho_0$, transports $\rho_0$ to $\rho_*$ in time $\ttot$. Specifically, the instantaneous distribution $\rho_t=\psi_t\#\rho_0 $ evolves according to 

\begin{equation}
\label{eqap:ot01}
    \partial_t \rho_t = -\nabla\left[u_t \rho_t\right],
\end{equation}

\noindent and one seeks to learn $u_t$ with the endpoint constraint $\rho_{\ttot} = \rho_*$. 
While initial CNF algorithms train on likelihood estimation, requiring multiple backward and forward ODE passes~\cite{chen_neural_2018}, the formulation of conditional flow matching (CFM) algorithms offer simulation free training objectives for $u_t$ without ODE solves. In this context, the general flow matching objective is given by
\begin{equation}
\label{eqap:ot1}
    \mathcal{L}_{\rm FM}\left[\theta\right] = \mathbb{E}_{t\sim \mathcal{U}(0,1),\rho_t(x)}||v^\theta(t,x)-u_t(x)||^2_2
\end{equation}
where $v^\theta$ regresses the sought after flow field. Of course, neither $u_t$ nor $\rho_t$ are known, and \eqref{eqap:ot1} is intractable as such. However these difficulties can be bypassed by introducing an alternative conditional flow matching (CFM) objective 
\begin{equation}
\label{eqap:ot2}
    \mathcal{L}_{\rm CFM}\left[\theta\right] = \mathbb{E}_{t\sim \mathcal{U}(0,1),\rho_t(x|z), q(z)}||v^\theta(t,x)-u_t(x|z)||^2_2
\end{equation}

\noindent where the latent variable $z$, sampled from some distribution $q(z)$, and the conditional distribution $\rho_t(x|z)$ are chosen such that

\begin{equation}
\label{eqap:ot3}
    \rho_t(x) = \int \rho_t(x|z)q(z)\der z,
\end{equation}

\noindent and $\rho_t(x|z)$ is generated by a conditional flow field $u_t(x|z)$ from initial datum $\rho_0(x|z)$. 
Crucially, both $\mathcal{L}_{\rm CFM}$ and $\mathcal{L}_{\rm FM}$ share the same minimizer. As a result, suitable choices of $z$ and $\rho_t(x|z)$ make the objective \eqref{eqap:ot2} explicit; in particular, the OTFM objective is obtained by considering the specific case

\begin{equation}
\label{eqap:ot4}
    \begin{alignedat}{1}
    &\rho_t(x|z) = \mathcal{N}\left[(1-t)x_0+t x_1,\sigma\right]\\
    &u_t(x|z) = (x_1-x_0)    
    \end{alignedat}
\end{equation}

\noindent where $x_0$ and $x_1$ are sampled from the optimal coupling $\pi^{\rm OT}$ between $\rho_0$ and $\rho_*$. Note that, by design, this OTFM formulation produces an OT pathway between $\rho_0$ and $\rho_*$ convolved with a Gaussian of width $\sigma$. However, such a formulation allow for easier score estimation, bypassing the gradient computation involved in Eq. (11) of the main text.

We now show how to numerically evaluate the score $s_t(x) \equiv \nabla \log \rho_t$ associated to the OTFM conditional pathway \eqref{eqap:ot4}. To simplify notations, we set $\ttot = 1$. The essential idea is to consider the OTFM framework in the light of the stochastic interpolant framework independently developed in \cite{albergo_building_2023}. Specifically, the conditional distribution
\begin{equation}
    \rho_t(x|z) = \mathcal{N}((1-t) x_0 + t x_1, \sigma)
\end{equation}
is the instantaneous conditional distribution of the stochastic interpolant 
\begin{equation}
 x_t = (1-t) x_0 + t x_1 + \sigma z, \hspace{10pt}z\sim \mathcal{N}(0,1),  \hspace{10pt}t\in [0,1], 
\end{equation}
conditioned on a given pair $(x_0,x_1)$ sampled from the optimal coupling $\pi^{\rm OT}$. Following the stochastic interpolant framework, the instantaneous distribution $\rho_t$ of $x_t$ obeys the following continuity equation
 \begin{equation}
     \partial_t \rho_t = -\nabla \left[u_t \rho_t\right]
 \end{equation}
 where $u_t$ is the unique minimizer of the objective function
 \begin{equation}
 \label{eq:appot}
     \mathcal{L}^u_{\rm SI}[\hat{u}] = \mathbbm{E}_{\left\{
     \substack{(x_0,x_1)\sim\pi^{\rm OT}, t\sim\mathcal{U}(0,1)\\
     z\sim \mathcal{N}(0,1), x =(1-t)x_0+tx_1+\sigma z}\right\}}||\hat{u}(x, t) - (x_1-x_0)||_2^2,
 \end{equation}
recovering the OTFM result \eqref{eqap:ot4}. Furthermore, the stochastic interpolant framework also allows for the explicit characterization of the score --- see equation (2.15) of \cite{albergo_building_2023} --- as the unique minimizer of

 \begin{equation}
     \mathcal{L}^s_{\rm SI}[\hat{s}] = \mathbbm{E}_{\left\{
     \substack{(x_0,x_1)\sim\pi^{\rm OT}, t\sim\mathcal{U}(0,1)\\
     z\sim \mathcal{N}(0,1), x =(1-t)x_0+tx_1+\sigma z}\right\}}||\hat{s}(x, t) + z\sigma^{-1}||_2^2,
 \end{equation}
which allows for a simpler objective evaluation.

 \section{Bimodal transform}

 In this section we provide the explicit functional forms of the source and target Gaussian Mixtures used in Fig. 3 of the main text:

 \begin{equation}
\label{eq:ot8}
\begin{alignedat}{1}
    &\rho_0(\xb) = \frac{1}{2}\left[p_1(\xb)+p_2(\xb)\right],\hspace{10pt}\rho_*(\xb) = \frac{1}{2}\left[p_3(\xb)+p_4(\xb)\right],\hspace{10pt}p_i\sim \mathcal{N}\left(\mub_i, \boldsymbol{\Sigma}_i\right)\\
    \\
    &\mub_1=\begin{bmatrix}
        6\\
        0
    \end{bmatrix}
    \hspace{10pt}
    \mub_2=\begin{bmatrix}
        -6\\
        0
    \end{bmatrix}
    \hspace{10pt}
    \mub_3=\begin{bmatrix}
        -10\\
        -10
    \end{bmatrix}
    \hspace{10pt}
    \mub_4=\begin{bmatrix}
        10\\
        10
    \end{bmatrix}\\
    \\
    &\boldsymbol{\Sigma}_1=\boldsymbol{\Sigma}_2=\mathbb{I}_2, \hspace{10pt}\boldsymbol{\Sigma}_3=\boldsymbol{\Sigma}_4=\mathbb{I}_2 \times 0.1.
\end{alignedat}
\end{equation}
Once the optimal controller is $b^*$ has been learned, the empirical mean dissipated heat is computed using the Stratonovitch rule as

\begin{equation}
    \avg{Q_{\ttot}} = \beta\avg{ \int_0^{\ttot}b(\Xb_t, t) \circ \der \Xb_t}.
\end{equation}
In Fig. 3, $t_{\rm tot}=1$, and $\beta=1$. For this figure, we have used the OTFM and score objectives stemming from the stochastic interpolant, as described in the preceding section.

\section{QSAP Dynamics}
\label{app:QSAP}

\begin{figure}[!ht]
    \centering
    \includegraphics[width=\linewidth]{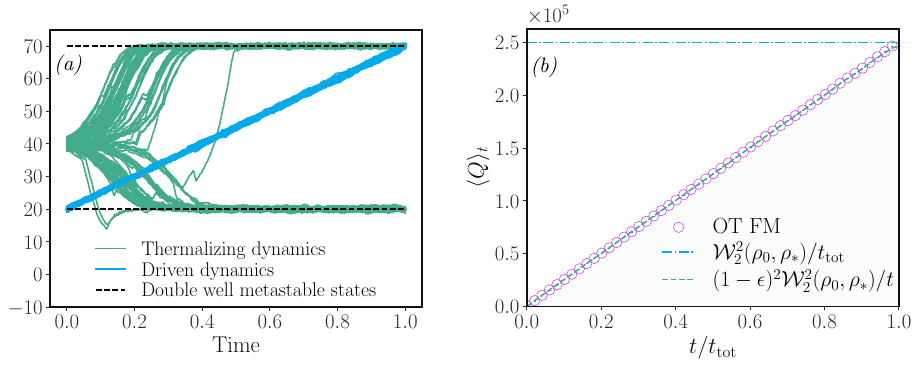}
    \captionsetup{
    font=footnotesize}
    \caption{\textbf{(a)} Thermalizing and driven dynamics for a cloud of $N=100$ QSAP with parameter values~$a=20,b=40,c=70,A=0.03$, $v_0=300,\ell=1,\lambda=10,\ttot=1, \beta=0.2$. Specific QSAP behavior is apparent in the thermalizing dynamics, with clear jagged trajectories corresponding to run and tumble type motion. \textbf{(b)} The observed heat at intermediate times from the ESA bound (green dashed line) compared with OTFM empirical measurements (circles). The Wasserstein speed limit is shown in blue.}
    \label{fig:qsap}
\end{figure}

As mentioned in the main text, we also directly employ control strategies based on OTFM to build minimum dissipation protocols in complex physical systems; we consider the problem of ``switching'' a system $\Xb_t$ of $N$ quorum sensing active particles (QSAP)~\cite{solon_generalized_2018} in a bistable potential $U$ from one metastable state to the other; the dynamics of this system obey the over-damped Langevin equation
\begin{equation}
\label{eq:qsap0}
    \der X^i_t = \left[-\nabla U(X^i_t)+ \eta^i(\Xb_t)\right]\der t+\sqrt{2\invt}\der W_t,
\end{equation}
in a metastable quartic and asymmetric potential $U$, such that the resulting force is given by
\begin{equation}
\label{eq:qsap1}
-\nabla U(X_t) = -A (X_t-a)(X_t-b)(X_t-c)
\end{equation}
with $a < c$ the respective positions of the two metastable states of the potential. The particles interact via the quorum sensing mechanism embedded in the interaction kernel $\eta$:
\begin{equation}
\label{eq:qsap2}
        \eta^i(\Xb_t) = \mathbbm{1}\left\{\underset{j\neq i}{\rm min}|X_t^i-X_t^j| > \ell \right\} \times v_0 \times \zeta^i_t
\end{equation}
where $\zeta^i_t$ is a telegraph process oscillating between $1$ and $-1$ with rate $\lambda$, $\ell$ is the interaction cutoff, and $v_0$ denotes the speed of a particle during an active phase. In words, the QSAP showcase a simple relaxation behavior when in close range of other particles, and adopt a run and tumble motion when isolated, as depicted in Fig.~\ref{fig:qsap}(a).

In the absence of any control, QSAPs settle in the metastable states, and transitions are enhanced by the activity, a dynamics we deem \textit{thermalizing}, in parallel to equilibrium dynamics in the same potential.
Although transition rates are increased by the run and tumble motion between metastable states, a collective transition remains unlikely. 
In the following we design a control force $\ctrl{b}(\Xb, t)$ to drive the interacting particle system from one well to the other in a finite time $\ttot$, and discuss the relevance of the ESA bound in that case.
Since the stationary distribution of the QSAP system is not known, we define the source and target distributions to be the Gaussian expansion of the activity-free equilibrium distribution near the well centers $x_i$,
\begin{equation}
\label{eq:qsap3}
    \rho_i(\Xb) \sim \mathcal{N}\left(\mathbf{1}x_i, \mathbb{I} \frac{\invt}{U''(x_i)}\right).
\end{equation}
As a result, the optimal control force $\ctrl{b} = \ctrl{\eta}+\ctrl{F}+\ctrl{v}$ decomposes in three terms: $\ctrl{\eta}(\Xb_t)$ cancels out the quorum sensing interaction; $\ctrl{F}(\Xb_t)$ cancels the conservative force and $\ctrl{v}(\Xb_t,t)$ maintains the instantaneous distribution on the Wasserstein geodesic between $\rho_0$ and $\rho_*$. 
For a given controller $b^{(c)}(\Xb_t, t)$ the equation of motion for a single particle then become
\begin{equation}
\label{eq:qsap4}
    \der X^i_t = \left[-\nabla U(X^i_t)+\eta^i(\Xb_t)+ b^{(c),i}(\Xb_t, t)\right]\der t +\sqrt{2\invt}\der B_t,
\end{equation}
and the mean dissipated heat is computed using the Stratonovitch rule as
\begin{equation}
    \avg{Q_{\ttot}} = \avg{\beta \int_0^{\ttot} \sum_{i=1}^{N} \left(\left[-\nabla U(X^i_t)+\eta^i(\Xb_t)+ b^{(c),i}(\Xb_t, t)\right] \circ \der X^i_t\right)}.
\end{equation}
Since perfect control may not be reached by physical controllers, we illustrate the accuracy gap by representing $\ctrl{\eta}$ and $\ctrl{F}$ as neural networks trained on thermalizing trajectory data.
Learned control forces perform particularly well, as the dynamical total dissipated heat is close to optimal. 
Fig.~\ref{fig:qsap} (b) shows the average dissipated heat along approximate control trajectories, including both the corresponding ESA bound and also the classical energy speed limit, highlighting the utility of focusing on the instantaneous distribution of the controlled particles. 
%

\section{Timescale separation - Slow driving regime - Non equilibrium case}

In this section, we provide details on the non equilibrium version of the timescale separation framework leading to the slow driving regime bound~(23) of the main text. 
For driven non-equilibrium systems described by the equations of motion

\begin{equation}
\begin{split}
    &\der \Xbc_t = b^{\rm c}(\Xbc_t, t)\der t\\
    & \der \Xb_t = -\epsilon^{-1}\left[\nabla U(\Xb_t - \Xbc_t) + f(\Xb_t - \Xbc_t)\right]\der t + \sqrt{2\beta^{-1}\epsilon^{-1}}\der \Wb_t, 
\end{split}
\label{app:timeneq1}
\end{equation}
with  $f$ a non conservative component, the argument is similar, but the expansion is done around the instantaneous stationary state 
\begin{equation}
    \rho^{\rm ss }_t(\xb) = e^{-\psi(\xb - \Xbc_t)} = \rho^{\rm ss}(\xb - \Xbc_t),
\end{equation}
where we have
\begin{equation}
    \nabla \left[(-\nabla U(\xb) + f(\xb))\rho^{\rm ss}(\xb) - \beta^{-1}\nabla \rho^{\rm ss}(\xb) \right] = 0 .
\end{equation}
Following the Hatano-Sasa definition of the housekeeping heat (see Eq. (15) in~\cite{hatano_steady-state_2001})
\begin{equation}
    \avg{\dot{Q}_{\rm hk}} = \mathbb{E}\left[(-\nabla U (\Xb_t)+f(\Xb_t) +\beta^{-1}\nabla \psi(\Xb_t))\circ \der \Xb_t\right]_{\rho^{\rm ss}}
\end{equation}
which is a constant in the slow driving regime and vanishing when $f = 0$, the dissipation conditioned on initial $\Xbc_0$ value is given by 

\begin{equation}
\avg{Q | \Xbc_0} = \ttot \avg{\dot{Q}_{\rm hk}} + \beta^{-1}\int_0^{\ttot} \der t \dot{\Xbc_t} \mathbb{E}\left[\nabla \psi(\xb - \Xbc_t)\right]_{\rho_t}
    \label{app:timeneq2}
\end{equation}
It is then clear that the second term on the r.h.s. can be expanded in a fashion similar to the equilibrium case, and leveraging the translational invariance along the pathway, we obtain

\begin{equation}
\avg{Q} \geq \ttot \avg{\dot{Q}_{\rm hk}} + C_{\rm neq} \frac{\was^2(\rho_0, \rho_*)}{\ttot}
    \label{app:timeneq3}
\end{equation}
where $C_{\rm neq}$ is defined as an autocorrelation function

\begin{equation}
    C_{\rm neq} = \beta^{-1}\int_0^\infty \der \tau \mathbb{E}\left[\nabla \psi(\Tilde{\Xb}_\tau) \nabla \psi(\Tilde{\Xb}_0)\right]_{\rho^{\rm ss}}
\end{equation}
over the modified dynamics
\begin{equation}
    \der \Tilde{\Xb}_\tau = \left[\nabla U(\Tilde{\Xb}_\tau) - f(\tilde{\Xb_\tau}) - 2 \beta^{-1} \nabla \psi(\tilde{\Xb}_\tau))\right]\der \tau + \sqrt{2\beta^{-1}}\der \Wb_\tau.
\end{equation}
Note that in contrast to the equilibrium case, these new dynamics are not strictly similar to the initial ones. In the case $\psi = \beta U$, we recover the equilibrium result (22) from the main text.

\section{Details on the driven dimer system}

In this section, we provide details on the driven dimer dynamics associated to Fig. 4 of the main text. The dimer comprises of two one dimensional particles $X_t^i$, $i \in \{1,2\}$ driven by an externally controlled particle $X_t^{\rm c}$. We wish to optimally drive $X_t^{\rm c}$ between source and target distributions defined respectively as a gaussian and a gaussian mixture~:

\begin{equation}
    \begin{split}
        &\rho_0 = \mathcal{N}(0,1)\\
        &\rho_* = \frac{1}{2}\left[\mathcal{N}(-10,0.5) + \mathcal{N}(10,0.5)\right]
    \end{split}
\end{equation}
In turn, the specific equations of motion we integrate are

\begin{equation}
    \begin{split}
        &\der X^{\rm c}_t = b^{c}(X_t^{\rm c}, t)\der t\\
        &\der X^{i}_t = \left[-\partial_x (U^{\rm dimer}(X_t^{i}-X_t^{j}) + U^{\rm c}(X_t^{i}-X_t^{\rm c}) + \eta^i(X_t^i - X_t^j)\right]\der t + \sqrt{2 \beta ^{-1}}\der W_t^i.
    \end{split}
    \label{app:dimer}
\end{equation}
We use a timestep $dt=10^{-4}$ over a duration $\ttot=100.0$, with $4.10^3$ samples and $\beta=1$. The specifics of the forces are~:
\begin{itemize}
    \item The interacting dimer potential is a quartic potential 
    \begin{equation}
        U^{\rm dimer}(r) = d_e\left(\frac{r^4}{4} - \frac{r^2}{2}\right)
    \end{equation}
    where $d_e = 5.0$.
    \item The coercive controller potential is a harmonic trap given by
    \begin{equation}
        U^{\rm c}(r) = - k_{\rm trap}\  r
    \end{equation}
    with $k_{\rm trap} = 3.0$.
    \item The active component $\eta$, which is disabled in the passive run, is the QSAP interaction term used in the QSAP cloud experiment described above
    \begin{equation}
        \eta^i(X_t^{i} - X_t^j) = \mathbbm{1}\left\{|X_t^i-X_t^j| > \ell \right\} \times v_0 \times \zeta^i_t
    \end{equation}
    with $\ell = 1.0, v_0 = 1.0$ and $\zeta^i_t$ is a telegraphic noise between $1$ and $-1$ with reversing rate $\gamma=0.1$.
    \item The driving field $b^{\rm c}$ is obtained using OTFM --- training and architecture details are provided in the training section.
, \end{itemize}
In the passive run, $C_{\rm dimer}$ is given by the force force autocorrelation function
\begin{equation}
    C_{\rm dimer} = \avg{\int_0^\infty \nabla(U_\tau^{\rm dimer} + U_\tau^{\rm c})^T \cdot \nabla(U_0^{\rm dimer} + U_0^{\rm c})\der \tau}
\end{equation}
which is numerically evaluated to $\simeq 2.0$ from equilibrium trajectories.

\section{Training}
\label{app:training}

In this section, we provide details on the various training architectures used throughout the main text. Table \ref{app:trainnig-table} recapitulates the values of the hyper-parameters we use for the training protocols. Below, we provide further information on training routines, in particular the dataset construction.

\subsection{OTFM - Fig. 3 of the main text}

To train the OTFM field $b^{\rm OT}$ on the objective \eqref{eq:appot}, we first generate $N$ independent samples $x_0^i$ and $x_1^i$ from $\rho_0$ and $\rho_*$ respectively. We then use the python \textbf{POT} optimal transport library to obtain empirical optimal coupling sample pairs $(x_0^{\rm OT},x_1^{\rm OT})_{1\leq i \leq N}$ which are used as inputs to carry out the OTFM optimization.

\subsection{QSAP - Fig. 1}

To control QSAP trajectories and illustrate the optimal driving procedure, we learn approximate representations of the external force $F^{(c)}\equiv - \nabla U$ and quorum sensing kernel $\eta^{(c)}\equiv \eta$ introduced in the main text. 

\begin{itemize}
    \item $\mathbf{F^{(c)}}$ --- Because the external force acts on each particle independently, we parameterize the network as $F^{(c)} : \mathbbm{R}^2 \to \mathbbm{R}$. We optimize the MSE loss between $F^{(c)}$ and the ground truth $\nabla U$ on independent samples $(t,x)_{1\leq i \leq N}$ generated from $t\sim \mathcal{U}(0,\ttot),\ x\sim \mathcal{N}(t, 2.0)$. We make this choice to help learn the control force in the vicinity of the desired controlled trajectory.   

    \item $\mathbf{\boldsymbol{\eta}^{(c)}}$ --- The quorum sensing kernel is a simple scaled Heaviside function. We parameterize the network as $\eta^{(c)} : \mathbbm{R} \to [0,1]$ and train on the MSE loss with the ground truth ${\rm Heavi}_\ell$, where $\ell$ is the QSAP cutoff parameter. The training data is drawn independently from a Normal distribution. Note that during inference, the control force is taken to be $\mathbbm{1}_{x>1/2}\circ \eta^{(c)}$ to recover the binary nature of the quorum sensing mechanism.
\end{itemize}

\subsection{Dimer driving - Fig. 4 of the main text}

The optimal controller $b^{\rm c}(x, t)$ involved in the dimer dynamics~\eqref{app:dimer} 
is obtained by the OTFM procedure described in the main text. Because this is a 1D system, optimal pairing of samples $(x_0, x_1)$ are produced by sorting the samples and pairing them in order.

\begin{table}[!ht]
\centering
\caption{Training protocol details}
\label{app:trainnig-table}
\resizebox{\columnwidth}{!}{%
\begin{tabular}{lllllll}
\hline
\multicolumn{1}{|l|}{}   & \multicolumn{1}{l|}{Network architecture}  & \multicolumn{1}{l|}{Optimizer}  & \multicolumn{1}{l|}{Scheduler} & \multicolumn{1}{l|}{Samples} & \multicolumn{1}{l|}{Batch Size} & \multicolumn{1}{l|}{Epochs} \\ \hline

\multicolumn{7}{c}{Fig. 3 of the main text}\\ \hline

\multicolumn{1}{|l|}{$b^{\rm OT}$}   & \multicolumn{1}{l|}{MLP with 1 hidden layer, 256 neurons and ReLU activation}   & \multicolumn{1}{l|}{Adam(lr=$3.10^{-4}$)}   & \multicolumn{1}{l|}{ReduceLROnPlateau(patience=5,factor=0.8,min\_lr=$10^{-6}$)} & \multicolumn{1}{l|}{$10^4$} & \multicolumn{1}{l|}{64} & \multicolumn{1}{l|}{$10^3$} \\\hline

\multicolumn{1}{|l|}{$b^{\times}$}   & \multicolumn{1}{l|}{MLP with 1 hidden layer, 256 neurons and ReLU activation}   & \multicolumn{1}{l|}{Adam(lr=$3.10^{-4}$)}   & \multicolumn{1}{l|}{ReduceLROnPlateau(patience=5,factor=0.8,min\_lr=$10^{-6}$)} & \multicolumn{1}{l|}{$10^5$} & \multicolumn{1}{l|}{64} & \multicolumn{1}{l|}{$5.10^3$} \\ \hline

\multicolumn{7}{c}{Fig. 1}\\ \hline

\multicolumn{1}{|l|}{$\eta^{(c)}$}   & \multicolumn{1}{l|}{MLP with 1 hidden layer, 512 neurons and ReLU activation}   & \multicolumn{1}{l|}{Adam(lr=$10^{-3}$)}   & \multicolumn{1}{l|}{ReduceLROnPlateau(patience=5,factor=0.8,min\_lr=$10^{-6}$)} & \multicolumn{1}{l|}{$10^4$} & \multicolumn{1}{l|}{64} & \multicolumn{1}{l|}{$10^3$} \\ \hline

\multicolumn{1}{|l|}{$F^{(c)}$}   & \multicolumn{1}{l|}{MLP with 1 hidden layer, 512 neurons and ReLU activation}   & \multicolumn{1}{l|}{Adam(lr=$10^{-3}$)}   & \multicolumn{1}{l|}{ReduceLROnPlateau(patience=5,factor=0.8,min\_lr=$10^{-6}$)} & \multicolumn{1}{l|}{$10^4$} & \multicolumn{1}{l|}{64} & \multicolumn{1}{l|}{$10^3$} \\ \hline

\multicolumn{7}{c}{Fig. 4 of the main text}\\ \hline

\multicolumn{1}{|l|}{$b^{\rm c}$}   & \multicolumn{1}{l|}{MLP with 1 hidden layer, 128 neurons and GeLU activation}   & \multicolumn{1}{l|}{Adam(lr=$10^{-4}$)}   & \multicolumn{1}{l|}{None} & \multicolumn{1}{l|}{$8.10^3$} & \multicolumn{1}{l|}{32} & \multicolumn{1}{l|}{$5.10^3$} \\ \hline

\end{tabular}
}
\end{table}

\newpage 
\bibliography{refs, references} 

\end{document}